\documentclass[prd,aps,nofootinbib,floatfix,10pt]{revtex4}
\usepackage{amsmath,graphicx,epsfig,amssymb}
\usepackage{inputenc}
\usepackage[usenames]{color}
\usepackage{ulem} 
\usepackage{bigstrut}
\usepackage{slashed}
\usepackage{multirow}
\usepackage{subfigure}

\allowdisplaybreaks

\begin{document}
	
\title{A global analysis of charmless two body hadronic decays\\ for anti-triplet charmed baryons}
\author{
Fei Huang$^{1}$, Zhi-Peng Xing$^{2}$\footnote{Corresponding author: zpxing@sjtu.edu.cn} and Xiao-Gang He$^{1,2,3}$  }
\affiliation{$^{1}$
                   INPAC, Key Laboratory for Particle Astrophysics and Cosmology (MOE), Shanghai Key Laboratory for Particle Physics and Cosmology,
                   School of Physics and Astronomy, Shanghai Jiao Tong University, Shanghai
		  200240, China
		 }		
\affiliation{$^{2}$  Tsung-Dao Lee Institute, Shanghai Jiao Tong University, Shanghai 200240, China}
\affiliation{$^{3}$
		   NCTS and Department of Physics, National Taiwan Universityt, Taipei 10617, Taiwan }

\begin{abstract}
Recently Belle collaboration reported new measurements for the branching fractions with the first observing two processes of $\mathcal{B}(\Xi_c^0\to\Lambda K^0_S)$,
$\mathcal{B}(\Xi_c^0\to\Sigma^0 K^0_S)$ and updating data for $\mathcal{B}(\Xi_c^0\to\Sigma^+ K^-)$. Combined with other known data on charmless two body decays of anti-triplet charmed baryons, a lot of information can be derived with the assistance of $SU(3)$ flavour symmetry. Using $SU(3)$ relations between different decay modes, we can give some predictions based on the new measurements which can be tested with the high luminosity experiments in the future. More interestingly, we find that a global fit is now possible with the addition of new Belle data. 
In general, there are 18 complex $SU(3)$ invariant amplitudes. We find that a scenario of all amplitudes being real can fit the data
well with a $\chi^2/d.o.f$ only  $0.773$. This  indicates that neglecting the phases of the amplitudes  is a reasonable assumption. When more data become available, one may be able to get more information for phases in the amplitudes.
 We give several comments on the feature of global fit regarding the branching fractions, relations between different decays, and decays involving $K^0$ and $\bar K^0$. Many of the unknown branching fractions and polarization asymmetry parameters of anti-triplet charmed baryon for charmless two body decays are predicted to be accessible by experiments at Belle, Belle~II, BES-III, and LHCb. The validity of $SU(3)$ for charmless two body hadronic decays can be more accurately tested.
\end{abstract}
	\maketitle
	
\section{introduction}
Charmless two body hadronic decays of anti-triplet charmed baryons have received considerable attentions both theoretically and experimentally~\cite{BESIII:2018ciw,BESIII:2018cvs,Belle:2020xku,Lyu:2021biq,He:2018joe,Geng:2019xbo,Zhao:2018mov,Wang:2017gxe,Chua:2018ikx,Hsiao:2020iwc}. 
A lot of data on these decays have been collected from experiments ~\cite{Belle:2018kzz,Belle:2021avh,BESIII:2020nme}. However, theoretical calculations for these decays have proven to be difficult due to the strong QCD interaction at the charm scale~\cite{Cheng:2021qpd}. The use of perturbative calculations is at the boundary of applicability.  It is believed that lattice QCD will ultimately provide reliable theoretical results\cite{Zhang:2021oja,Meinel:2017ggx,Bahtiyar:2021voz}. But it is still far from reaching that goal. Many of the theoretical studies for these decays rely on modeling QCD dynamics for predictions and to test different models\cite{Aliev:2021wat,Zhao:2018zcb,Zhao:2021sje,Liu:2010bh,Cheng:1991sn,Li:2016qai,Guo:2005qa}. One of the most used models is based on $SU(3)$ flavour symmetry which parametrize amplitudes according to symmetry properties~\cite{Savage:1989qr,Savage:1991wu,Sharma:1996sc,Cheng:2014rfa,Geng:2018plk,Geng:2019bfz,He:2015fwa,Hsiao:2021nsc,Lu:2016ogy,Huang:2021jxt,Wang:2017azm,Shi:2017dto}. Although one cannot obtain the absolute size of the decay amplitudes based on the first principle calculations, relations can be obtained among different decay modes to test the model. When enough  experimental data become available, all the amplitudes classified by $SU(3)$ symmetry properties can be determined and then systematic predictions can be made to further test the model to provide more information for strong QCD interactions at the charm scale~\cite{He:2018joe,Geng:2019xbo}. In this paper,  we carry out an $SU(3)$ analysis for charmless two body hadronic decays of anti-triplet charmed baryons with existing data and combine the
recently measured new decay branching fractions from Belle collaboration~\cite{Belle:2021avh},
\begin{eqnarray}
&&\mathcal{B}(\Xi_c^0\to\Lambda  K^0_S)=(4.12\pm0.14\pm0.21\pm1.19)\times 10^{-3},\notag\\
&&\mathcal{B}(\Xi_c^0\to\Sigma^0  K^0_S)=(0.69\pm0.10\pm0.08\pm0.20)\times 10^{-3},\notag\\
&&\mathcal{B}(\Xi_c^0\to\Sigma^+ K^-)=(2.21\pm0.13\pm0.19\pm0.64)\times 10^{-3}\;.\label{eq1}
\end{eqnarray}

$SU(3)$ symmetry predicts relations between different decays and it can be used to test the model. 
Existing data show that the predicted relation $\Gamma(\Lambda^{+}_{c}\to\Sigma^{0}\pi^{+})= \Gamma(\Lambda^{+}_{c}\to\Sigma^{+}\pi^{0})$ holds well, which gives some confidence for the model. It is tempting to make a more elaborated analysis using new data in combination with existing data to test other possible relations predicted by $SU(3)$. 
It sparked our interest in systematic $SU(3)$ analysis of charmless two body hadronic decays of anti-triplet charmed baryons. 
In this work, we carry out such an analysis. We find that $SU(3)$ flavor symmetry can describe data well, unlike the case in semileptonic anti-triplet charmed baryon decays~\cite{He:2021qnc}. Tests of $SU(3)$ symmetry not only using certain relations but in wider scope also become possible. 
 In general there are 18 complexes $SU(3)$ invariant amplitudes for charmless two body hadronic decays of anti-triplet charmed baryons. We find the subset of all amplitudes is real can  fit the data well with a $\chi^2/d.o.f$ only  $0.773$. This indicates that neglecting the phases of the amplitudes  is a reasonable assumption. When more data become available, one may be able to get more information for phases in the amplitudes. We await more data to come to do such an analysis.
 We give several comments on the features of global fit regarding the branching fractions, relations between different decays and decays involving $K^0$ and $\bar K^0$. Many of the unknown branching fractions and asymmetry parameters are predicted to be accessible by experiments at Belle, Belle~II, BESIII, and LHCb. The validity of $SU(3)$ for charmless two body hadronic decays can be more accurately tested.

The paper is organized as follows. In Sec.~\ref{sec1}, we give the theoretical framework of $SU(3)$ symmetry analysis using the irreducible representation amplitude (IRA) method.  In Sec.~\ref{sec2}, we discuss possible relations among different decay modes by $SU(3)$ symmetry to see how predictions for other decays can be obtained. In Sec.~\ref{sec3}, we use experimental data to carry out a global fit to obtain $SU(3)$ irreducible amplitudes and to predict some unknown branching fractions for charmless two body hadronic decays of anti-triplet charmed baryons. In Sec.~\ref{sec4}, the $K_{S}^{0}-K_{L}^{0}$ mixing effects on some of the decay modes are discussed. A brief conclusion will be given in the last section.

\section{$SU(3)$ symmetry decay amplitudes}\label{sec1}

The decays we study are one of the charmed baryons in anti-triplet $T_{c\bar 3}$  to a baryon in the ground octet $T_8$ plus a meson in the nonet pesudoscalar $P$. The component states in representations are written as
\begin{eqnarray}\
T_{c\bar3}=
\begin{pmatrix}
0& \Lambda_c^+ &\Xi_c^+ \\
-\Lambda_c^+ & 0&\Xi_c^0\\
-\Xi_c^+& -\Xi_c^0&0
\end{pmatrix},\quad
T_8&=&
\begin{pmatrix}
\frac{\Sigma^0}{\sqrt{2}}+\frac{\Lambda}{\sqrt{6}}& \Sigma^+ &p \\
\Sigma^- & -\frac{\Sigma^0}{\sqrt{2}}+\frac{\Lambda}{\sqrt{6}}&n\\
\Xi^-& \Xi^0&-\frac{2\Lambda}{\sqrt{6}}
\end{pmatrix},\quad
P=
\begin{pmatrix}
\frac{\pi^0+\eta_q}{\sqrt{2}}& \pi^+ &K^+ \\
\pi^- & \frac{-\pi^0+\eta_q}{\sqrt{2}}&K^0\\
K^-& \bar{K}^0&\eta_s
\end{pmatrix}.
\end{eqnarray}
Here the $SU(3)$ flavour anti-triplet charmed baryons can be also represented as $(T_{c\bar 3})_i=\epsilon_{ijk}(T_{c\bar 3})^{[jk]}=(\Xi_c^0,-\Xi_c^+,\Lambda^+_c)$.
The $\eta_s$ and $\eta_q$ are the mixture of meson singlet $\eta_1$ and octet $\eta_8$ components:
\begin{eqnarray}
\eta_8=\frac{1}{\sqrt{3}}\eta_q-\sqrt{\frac{2}{3}}\eta_s,\quad \eta_1=\sqrt{\frac{2}{3}}\eta_q+\frac{1}{\sqrt{3}}\eta_s.
\end{eqnarray}
Moreover, the physical $\eta$ and $\eta'$ states are mixtures of  $\eta_q={1\over \sqrt{2}}(u\bar u+d\bar d)$ and $\eta_s=s\bar s$ given by 
\begin{eqnarray}
\begin{pmatrix}
\eta \\
\eta^\prime\\
\end{pmatrix}=\begin{pmatrix}
\cos\phi&-\sin\phi \\
\sin\phi&\cos\phi\\
\end{pmatrix}\begin{pmatrix}
\eta_q \\
\eta_s\\
\end{pmatrix},
\end{eqnarray}
with $\phi=(39.3\pm1.0)^{\circ}$~\cite{Zyla:2020zbs}.

For the decay processes we are discussing, the effective Hamiltonian in the standard model (SM) can be divided into three groups: Cabibbo-allowed,  Cabibbo-suppressed, and doubly Cabibbo-suppressed,

\begin{eqnarray}
\mathcal{H}_{eff}=\sum_{i=1,2}\frac{G_F}{\sqrt{2}}C_i(V_{cs}V^*_{ud}O^{s\bar d u}_i+V_{cq}V^*_{uq}O^{q \bar qu}_{i}+V_{cd}V^*_{us}O_i^{d\bar s u})+h.c. ,\quad q=s,d,
\end{eqnarray}
with
\begin{eqnarray}
O^{q_1\bar q_2 q_3}_1=(\bar{q_1}_\alpha c_\beta)_{V-A}(\bar{q_3}_\beta {q_2}_{\alpha})_{V-A},\quad   O^{q_1\bar q_2 q_3}_2=(\bar{q_1}_\alpha c_\alpha)_{V-A}(\bar{q_3}_\beta {q_2}_{\beta})_{V-A}.
\end{eqnarray}
The tree operators transform under the $SU(3)$ flavour symmetry  as: $3\otimes \bar 3\otimes 3=3\oplus3\oplus\bar6\oplus15$.
The Cabibbo allowed $c\to s \bar d u$ transition is proportional to $V^*_{cs}V_{ud} \approx 1$,  the non-zero entries of the Hamiltonians  under $SU(3)$ are
\begin{eqnarray}
(H_{\bar 6})^{31}_2=-(H_{\bar 6})^{13}_2=1,\quad (H_{15})^{31}_2=(H_{15})^{13}_2=1.
\end{eqnarray}
The Cabibbo-suppressed $c\to u\bar ss(\bar dd)$  transitions are proportional to  $V^*_{cs} V_{us}(-V^*_{cd} V_{ud}) \approx \sin\theta = 0.2265\pm0.00048$\cite{Zyla:2020zbs}, one has
\begin{eqnarray}
&&(H_{\bar 6})^{31}_3=-(H_{\bar 6})^{13}_3=(H_{\bar 6})^{12}_2=-(H_{\bar 6})^{21}_2=\sin\theta,\notag\\&& (H_{15})^{31}_3=(H_{15})^{13}_3=-(H_{15})^{21}_2=-(H_{15})^{12}_2=\sin\theta.
\end{eqnarray}
The doubly Cabibbo-suppressed $c\to d\bar s u $ transitions are proportional to $V^*_{cd}V_{us}\approx \sin^2\theta$,
\begin{eqnarray}
(H_{\bar 6})^{21}_3=-(H_{\bar 6})^{12}_3=\sin^2\theta,\quad (H_{15})^{21}_3=(H_{15})^{12}_3=\sin^2\theta.
\end{eqnarray}
 In the above, we have used the Hamiltonians for Cabibbo allowed and doubly suppressed decays without the $H_{3}$ representation. We also used  the relation of $V_{cd}V^*_{ud}=-V_{cs}V^*_{us}-V_{cb}V^*_{ub}\approx -V_{cs}V^*_{us}$, which leads to cancellation $H_{3}$ contributions of the two Cabibbo suppressed terms to high accuracy. At the loop level, a small $H_{3}$ effect will be generated, but can also be neglected~\cite{He:2018joe}.
Using the anti-symmetric tensor $\epsilon_{kmi}$ to contract $(H_{\bar 6})^{km}_j$,  we can define a new form for the $H_{\bar 6}$ representation, 
\begin{eqnarray}\
(H_{\bar 6})_{\{ij\}}=\frac{1}{2}\epsilon_{kmi}(H_{\bar 6})^{km}_j=
\begin{pmatrix}
0&0&0 \\
0& 1& \sin\theta\\
0& \sin\theta&-\sin^2\theta
\end{pmatrix}.
\end{eqnarray}
The amplitudes of processes $T_{c\bar3}\to T_8 P$ in irreducible representation amplitude (IRA) can be written as 
\begin{eqnarray}
\mathcal{M}&=&a_{15}(T_{c\bar{3}})_i(H_{\overline{15}})^{\{ik\}}_j(\overline{T_8})^j_kP^l_l+b_{15}(T_{c\bar{3}})_i(H_{\overline{15}})^{\{ik\}}_j(\overline{T_8})^l_kP^j_l\notag\\
&+&c_{15}(T_{c\bar{3}})_i(H_{\overline{15}})^{\{ik\}}_j(\overline{T_8})^j_lP^l_k+d_{15}(T_{c\bar{3}})_i(H_{\overline{15}})^{\{jk\}}_l(\overline{T_8})^l_jP^i_k\notag\\
&+&e_{15}(T_{c\bar{3}})_i(H_{\overline{15}})^{\{jk\}}_l(\overline{T_8})^i_jP^l_k+a_{6}(T_{c\bar{3}})^{[ik]}(H_{\overline{6}})_{\{ij\}}
(\overline{T_8})^j_kP^l_l\notag\\&+&b_{6}(T_{c\bar{3}})^{[ik]}(H_{\overline{6}})_{\{ij\}}(\overline{T_8})^l_kP^j_l+c_{6}(T_{c\bar{3}})^{[ik]}(H_{\overline{6}})_{\{ij\}}(\overline{T_8})^j_lP^l_k\notag\\&+&d_{6}(T_{c\bar{3}})^{[kl]}(H_{\overline{6}})_{\{ij\}}(\overline{T_8})^i_kP^j_l.
 \label{su3}
\end{eqnarray}
By expanding out the above expression, we obtain $SU(3)$ amplitudes for the charmless two body hadronic decay processes of $T_{c\bar3}\to T_8 P$  in Table~\ref{table1} and Table~\ref{tableeta}.

\section{Amplitude relations predicted by $SU(3)$ symmetry}\label{sec2}

$SU(3)$ symmetry predicts relations among different decay modes which can be used to predict some not yet known branching fractions and be tested. We discuss how can such analysis can be carried in this section.
From Table~\ref{table1}, we obtain the following amplitude relations among difference decay modes,
\begin{eqnarray}
&&\mathcal{M}(\Lambda^{+}_{c}\to\Sigma^{0}\pi^{+})= - \mathcal{M}(\Lambda^{+}_{c}\to\Sigma^{+}\pi^{0}),\notag\\
&& \mathcal{M}(\Lambda^{+}_{c}\to\Sigma^{+}K^{0})= { }\mathcal{M}(\Xi^{+}_{c}\to p\bar{K}^{0}),\notag\\
&& \mathcal{M}(\Lambda^{+}_{c}\to n\pi^{+})= { }\mathcal{M}(\Xi^{+}_{c}\to\Xi^{0}K^{+}),\notag\\
&& \mathcal{M}(\Xi^{0}_{c}\to n\bar{K}^{0})= -\mathcal{M}(\Xi^{0}_{c}\to\Xi^{0}K^{0}),\notag\\
&& \mathcal{M}(\Lambda^{+}_{c}\to nK^{+})=-\sin^2\theta\mathcal{M}(\Xi^{+}_{c}\to\Xi^{0}\pi^{+}), \notag\\
&& \mathcal{M}(\Xi^{+}_{c}\to n\pi^{+})= -\sin^2\theta\mathcal{M}(\Lambda^{+}_{c}\to\Xi^{0}K^{+}),\notag\\
&& \mathcal{M}(\Xi^{+}_{c}\to\Sigma^{+}K^{0})=- \sin^2\theta \mathcal{M}(\Lambda^{+}_{c}\to p\bar{K}^{0}),\notag\\
&& \mathcal{M}(\Xi^{+}_{c}\to\Sigma^{+}\bar{K}^{0})= -\frac{1}{\sin^2\theta} \mathcal{M}(\Lambda^{+}_{c}\to pK^{0}),\notag\\
&& \mathcal{M}(\Xi^{0}_{c}\to\Sigma^{-}K^{+})=-\sin\theta\mathcal{M}(\Xi^{0}_{c}\to\Sigma^{-}\pi^{+})=\sin\theta\mathcal{M}(\Xi^{0}_{c}\to\Xi^{-}K^{+})= \sin^2\theta\mathcal{M}(\Xi^{0}_{c}\to\Xi^{-}\pi^{+}),\notag\\
&&\mathcal{M}(\Xi_c^0\to p\pi^-)={\sin\theta}\mathcal{M}(\Xi_c^0\to pK^-)=-{\sin\theta}\mathcal{M}(\Xi_c^0\to \Sigma^+\pi^-)=\sin^2\theta \mathcal{M}(\Xi_c^0\to \Sigma^+K^-). 
\label{relations}
\end{eqnarray}

 Using available data for $\Gamma(\Lambda^{+}_{c}\to\Sigma^{0}\pi^{+})$ and $\Gamma(\Lambda^{+}_{c}\to\Sigma^{+}\pi^{0})$ decays, the first equation above would predict
$\Gamma(\Lambda^{+}_{c}\to\Sigma^{0}\pi^{+})= { }\Gamma(\Lambda^{+}_{c}\to\Sigma^{+}\pi^{0})$. Using data  in Table~\ref{exp1} , the relation predicted is well respected.  It is tempting to use the known experimental data and  Eq.(\ref{relations}) to predict other branching fractions.
The newly measured experimental data $\mathcal{B}(\Xi_c^0\to\Sigma^+ K^-)$ in Eq.(\ref{eq1}) by Belle will predict the branching fractions of $\mathcal{B}(\Xi_c^0\to p \pi^-)=(5.817\pm 1.79)\times 10^{-6}$, and 
$\mathcal{B}(\Xi_c^0\to p K^-)=\mathcal{B}(\Xi_c^0\to \Sigma^+ \pi^-)=(1.113\pm 0.349)\times 10^{-4}$. These relation can be used to test the model. However, we find that the $SU(3)$ relation between Cabibbo allowed process $\Xi^0_c \to \Xi^- \pi^+$ and Cabibbo suppressed $\Xi^0_c \to \Xi^- K^+$ in Eq.(\ref{relations}) and using experimental data $\mathcal{B}(\Xi^0_c \to \Xi^- \pi^+)$, with exact $SU(3)$ symmetry the branching fraction for  $\Xi^0_c \to \Xi^- K^+$ is predicted to $(7.3 \pm 1.6)\times 10^{-4}$, which show deviation from the central value of  the data $(3.9\pm 1.2)\times 10^{-4}$.  This cast doubt on the validity of $SU(3)$ symmetry.

Whether the different central values are due to $SU(3)$ breaking or errors in data can be clarified with more data.
The usage of relations predicted should also be more carefully examined. We note that there are two reasons for relation $\Gamma(\Lambda^{+}_{c}\to\Sigma^{0}\pi^{+}) = \Gamma(\Lambda^{+}_{c}\to\Sigma^{+}\pi^{0})$ to hold. One is that this relationship is actually a result of isospin symmetry which is known to be better respected for the more relations involve s-quark in the process.
Another reason is that the phase space for  $\Lambda^{+}_{c}\to\Sigma^{0}\pi^{+}$ and $\Lambda^{+}_{c}\to\Sigma^{+}\pi^{0}$ is almost the same. But for $\Xi^0_c \to \Xi^- \pi^+$ and $\Xi^0_c \to \Xi^- K^+$, the phase spaces are different and also involve s-quark in the decays. The agreements would be better with $\mathcal{B}(\Xi^0_c \to \Xi^- K^+)=(6.62\pm1.48)\times 10^{-4}$ after using actual particle masses. As will be discussed later, there are actually two amplitudes, the scalar and pseudoscalar ones, for charmless two body hadronic decays of anti-triplet charmed baryons, the relative size of them also sensitively depends on the actual particle masses. When taken into account in a more correlated way, the branching fraction is predicted to be $(4.70\pm 0.08)\times10^{-4}$ in Table~\ref{table1} which is much closer to data. All the other relations in Eq.(\ref{relations}) should be viewed with modifications mentioned when obtaining respective branching fractions. Therefore a better way of treating the relations would be taking the amplitudes relations for the decay amplitudes, but using the masses of the actual particles for the phase spaces for the decays. This way of treating relations predicted will also ease the potential problem from the asymmetry parameters $\alpha$ for different decays in Eq.(\ref{relations}) which will be discussed in the next section. In exact $SU(3)$ symmetry, a similar analysis would show that the asymmetry parameters $\alpha$ for decay channels in Eq.(\ref{relations}) are all equal. We find that the values for $\alpha$ are also very sensitive to the actual hadron masses of the decays.  We, therefore, think that using actual masses for each process would give a more realistic prediction which partially takes into some obvious $SU(3)$ breaking effects for particle masses into account. In our later analysis, we will follow the above procedure.

There are several relations involving $K^0$ or $\bar K^0$ in the final states. These data need to be careful interpreted. For example, the measured branching fraction $(1.59\pm0.08)\%$ for  $\mathcal{B}(\Lambda^{+}_{c}\to p K_S^0)$  is for the final state $K^0_S$. But it is related to $K^0$ and $\bar K^0$ by  
\begin{eqnarray}
K_S^0=\frac{1}{\sqrt{2}}(K^0-\bar{K}^0),\quad K_L^0=\frac{1}{\sqrt{2}}(K^0+\bar{K}^0),
\end{eqnarray}
where small CP violating effect has been neglected. Therefore one also predict related decay invovling $K^0_L$ in the final state.
The following $SU(3)$ relations can be derived 
\begin{eqnarray}
&& \mathcal{M}(\Xi^{+}_{c}\to\Sigma^{+}K^{0})=- \sin^2\theta\mathcal{M}(\Lambda^{+}_{c}\to p\bar{K}^{0}),\notag\\&& \mathcal{M}(\Xi^{+}_{c}\to\Sigma^{+}\bar{K}^{0})= -\frac{1}{\sin^2\theta}\mathcal{M}(\Lambda^{+}_{c}\to pK^{0})\;.
\end{eqnarray}
The above would imply the relations for $K^0_L$ and $K^0_S$ in the final states,
\begin{eqnarray}
&& \mathcal{M}(\Lambda^{+}_{c}\to pK_S^{0})=\frac{\mathcal{M}(\Xi^{+}_{c}\to\Sigma^{+}K_L^{0})+\mathcal{M}(\Xi^{+}_{c}\to\Sigma^{+}K_S^{0})}{2\sin^2\theta}+\frac{\sin^2\theta}{2}(\mathcal{M}(\Xi^{+}_{c}\to\Sigma^{+}K_S^{0})-\mathcal{M}(\Xi^{+}_{c}\to\Sigma^{+}K_L^{0})),\notag\\
&& \mathcal{M}(\Lambda^{+}_{c}\to pK_L^{0})=-\frac{\mathcal{M}(\Xi^{+}_{c}\to\Sigma^{+}K_L^{0})+\mathcal{M}(\Xi^{+}_{c}\to\Sigma^{+}K_S^{0})}{2\sin^2\theta}+\frac{\sin^2\theta}{2}(\mathcal{M}(\Xi^{+}_{c}\to\Sigma^{+}K_S^{0})-\mathcal{M}(\Xi^{+}_{c}\to\Sigma^{+}K_L^{0})).\notag\\
\end{eqnarray}
It can be seen that the amplitude $\mathcal{M}(\Xi^{+}_{c}\to\Sigma^{+}K_L^{0})=-\mathcal{M}(\Xi^{+}_{c}\to\Sigma^{+}K_S^{0})$ in the leading order of $\sin^2\theta$. In this order, we obtain $\mathcal{B}(\Lambda^{+}_{c}\to p K_L^0)=(1.59\pm0.08)\%$ . But the convergence of this expansion will be violated if $(\mathcal{M}(\Xi^{+}_{c}\to\Sigma^{+}K_L^{0})+\mathcal{M}(\Xi^{+}_{c}\to\Sigma^{+}K_S^{0}))$ is in order of $\mathcal O(\sin^4\theta)$.
Therefore these interesting relations can give precise predictions only if at least one of the channels $\mathcal{B}(\Xi^{+}_{c}\to\Sigma^{+}K_S^{0})$, $\mathcal{B}(\Xi^{+}_{c}\to\Sigma^{+}K_L^{0})$ and $ \mathcal{B}(\Lambda^{+}_{c}\to pK_L^{0})$ is measured.
More measurements of this channel will provide further information about $SU(3)$ relations.

There are also several relations, 
\begin{eqnarray}
    && \mathcal{M}(\Lambda^{+}_{c}\to\Sigma^{+}K^{0})= { }\mathcal{M}(\Xi^{+}_{c}\to p\bar{K}^{0}),\notag\\&& \mathcal{M}(\Lambda^{+}_{c}\to n\pi^{+})= { }\mathcal{M}(\Xi^{+}_{c}\to\Xi^{0}K^{+}),\notag\\&& \mathcal{M}(\Xi^{0}_{c}\to n\bar{K}^{0})= { }\mathcal{M}(\Xi^{0}_{c}\to\Xi^{0}K^{0}),
\end{eqnarray}
whose branching fractions have  not been measured on both sides of the relation. We hope some of these branching fractions can be measured to  predict other branching fractions to test the model.  Note that  for the processes involving final state $K^0$ and $\bar{K}^0$, we can not give any predictions until both final states involving $K^0_S$ and $K^0_L$ are measured in the experiment. We eagerly hope more experimental data will be available in the future to allow us to give more meaningful comments.

\section{Determination of $SU(3)$ amplitudes by a global data fitting}\label{sec3}

In this section, we discuss a global analysis of charmless two body hadronic decays of anti-triplet charmed baryons based on $SU(3)$ symmetry. To carry out  such an analysis we should have enough data points to have some handle on the $SU(3)$ irreducible amplitudes. From Eq.(\ref{su3}), we see that there are 9 such amplitudes. In fact, each of these amplitudes should be considered to have both the combination of scalar and pseudoscalar form factors.
Generically, we can express them as
\begin{eqnarray}
  q_6&=&G_F\bar{u}(f^q_6 - g^q_6\gamma_5)u,\quad q=a,b,c,d,\notag\\
  q_{15}&=&G_F\bar{u}(f^q_{15} - g^q_{15}\gamma_5)u,\quad q=a,b,c,d,e,\label{ff}
\end{eqnarray}
according to the $(V- A)$ current nature of the effective Hamiltonian due to W exchange. 
In the above equations, each of these amplitudes has scalar and pseudoscalar form factors corresponding to the parity violating and conserving amplitude respectively. Therefore there are total 18 amplitudes.

Including the two newly measured branching fractions by Belle,  we see that there are 16 branching fractions in Table~\ref{exp1} that have been measured.  It is not possible to make a meaningful global data fitting if we only consider the 16 branching fractions. 
In refs. \cite{Geng:2019xbo,Zhao:2018mov,Wang:2017gxe}  under the approximation neglecting a
 part of amplitudes due to $H_{15}$ or omitting some amplitudes, global fitting was carried out. 
 However, several predicted branching fractions do not agree with data well as can be seen in Table~\ref{exp1}. For example, the branching fraction of process $\Xi_c^0\to\Sigma^+ K^-$ is measured as $0.221\pm0.068\%$, which has more than $5\sigma$ deviation in previous works \cite{Geng:2019xbo,Zhao:2018mov} and $4\sigma$ deviation in \cite{Wang:2017gxe}.
Therefore a global fit keeping all amplitudes is necessary for a more realistic study. 
To this end, we notice that polarization parameter $\alpha$ for 5 of the decays have been measured. They provide additional constraints. If these amplitudes are real, the number of parameters of 18 is less than the total data points of 21 making a global fit possible.  Therefore, we  take this as a scenario to do our analysis and a more general analysis will be available with more and more experimental data is coming in the future. Whether this scenario is a good one to work with will be judged by the fitting results. We find the subset of all amplitudes is real can  fit the data well with a $\chi^2/d.o.f$ only  $0.773$.  This gives us confidence in the results obtained. This will be our working assumption in our following analysis. 

For the process of anti-triplet charmed baryon decays $B_c\to B_n M$, the polarization angular distribution of decay width is easily written as
\begin{eqnarray}
\frac{d\Gamma}{d\cos\theta_M}=\frac{G_{F}^{2}|\vec{p}_{B_{n}}|(E_{B_n}+M_{B_{n}})}{8\pi M_{B_c}}(|F|^2+\kappa^2 |G|^2)(1+\alpha \hat\omega_i\cdot\hat p_{B_n} ),\label{decaywidth}
\end{eqnarray}
where $\hat\omega_i$ and $\hat p_{B_n}$ are the unit vector of initial state spin and final state momentum, respectively. Depending on the specific processes, the $F$ and $G$ linear functions of $f_i$ and $g_i$ are the scalar and peseudoscalar form factors, respectively.  The parameter $\alpha$~\cite{He:2015fsa} is given by
\begin{eqnarray}
\alpha=2\rm{Re}(F*G)\kappa/(|F|^2+\kappa^2 |G|^2), \kappa=|\Vec{p}_{B_n}|/(E_{B_n}+M_{B_n}).
\end{eqnarray}
 With data from measurements of $\alpha$ one can constraint separately the $G$ and $F$ form factors. 
The parameter $\kappa$ writing in terms of masses is given by
\begin{eqnarray}
\kappa^2 = {(M_{B_c} - M_{B_n})^2 - M^2_M \over (M_{B_c} + M_{B_n})^2 - M^2_M}\;.
\end{eqnarray}
Here $M_{B_c}$, $M_{B_n}$, and $M_M$ are the masses of anti-triplet charmed baryons, octet baryons, and mesons, respectively. This introduces additional sensitivity to relation predictions for each decay process mentioned before.

Before carrying out our global fitting,  we point out a special feature for the $SU(3)$ amplitudes that the $SU(3)$ amplitudes $a_6$, $a_{15}$ only contribute to the processes with final state $\eta$ and $\eta^\prime$ and the corresponding relations and data shown in Table~\ref{tableeta}. 
The $a_6$ and $a_{15}$ actually have four form factors $f^a_6$, $g^a_6$, $f^a_{15}$ and $g^a_{15}$. Experiments only  $\Lambda^{+}_{c}\to p\eta$, $\Lambda^{+}_{c}\to \Sigma^{+}\eta$ and $\Lambda^{+}_{c}\to \Sigma^{+}\eta^\prime$ three decay modes have been measured. This shows that completely determining the $SU(3)$ decay amplitudes is not possible. But we note that the three measured processes involve only $a_6 - a_{15}$, therefore we can do a meaningful global fit to include them to predict branching fractions not yet measured whose decay amplitudes only depend on $a_6-a_{15}$. 
For this purpose we define
\begin{eqnarray}
f^a=f^a_6-f^a_{15},\quad f^{a\prime}=f^a_6+f^a_{15},\quad g^a=g^a_6-g^a_{15},\quad g^{a\prime}=g^a_6+g^a_{15},
\end{eqnarray}
which corresponds to the new amplitude
\begin{eqnarray}
a=a_6-a_{15},\quad a^\prime=a_6+a_{15}.
\end{eqnarray}
Since the three experimental data which involve $\eta$ and $\eta^\prime$ only  rely on the form factors $f^a$ and $g^a$, 
therefore processes depend on $f^{a\prime}$ and $g^{a\prime}$ cannot be determined from our fit which are indicated by ``$-$'' in Table \ref{tableeta}. In our global fit, we have 21 data points, and 16 parameters,
the $f$ and $g$ amplitudes in $(a,\; b_6,\; c_6,\; d_6,\; b_{15},\;c_{15},\;d_{15},\;e_{15}$). Assuming these amplitudes are all real, we can 
carry out a meaningful global fit.
More experimental data are needed to predict the processes involving $a^\prime$.

 Using known experimental data on branching fractions and polarization asymmetry parameters $\alpha$ in Table~\ref{exp1}, we have performed our global fitting. We use the  Nonlinear Least Squares Fitting (lsqfit) package in reference~\cite{Peter:2020} for our analysis. With its definition of statistical data in the package, we get the $\chi^{2}/d.o.f=$  0.744 results, which indicates that the fit is a reasonable one. We conclude that $SU(3)$ symmetry gives a reasonable description of charmless two body hadronic decays of anti-triplet charmed baryons. The predictions are shown in Tables~\ref{table1} and \ref{tableeta} can be used to further test $SU(3)$ symmetry for charmless two body hadronic decays of anti-triplet charmed baryons.
 
As mentioned earlier that the amplitude $a_6+a_{15}$ corresponding to the form factors $f^{a\prime}$ and $g^{a\prime}$ cannot be determined in our fit without new inputs of  experimental data involve $a_6+a_{15}$.
Therefore new experimental data on this type of charmless two body anti-triplet charmed baryons having $\eta$ and $\eta^\prime$ in the final states is crucial for a complete test of $SU(3)$ symmetry. We hope in the near future some of these decays can be measured experimentally.

\section{$K_{S}^{0}-K_{L}^{0}$ asymmetries }\label{sec4}

In this section, we study asymmetry between a  $\mathcal{B_{C}} \to \mathcal{B}_n K_{S}^{0}$ and $\mathcal{B_{C}} \to \mathcal{B}_n K_{L}^{0}$ decay. In this type of decays, due to relations in Eq.(\ref{relations}),  for some of the decays there are interference between the Cabibbo-favored and doubly Cabibbo-suppressed amplitudes the asymmetry defined below does not vanish~\cite{Wang:2017gxe} \begin{align}
R(\mathcal{B}_{C}\rightarrow\mathcal{B} K_{S,L}^{0})=\frac{\Gamma(\mathcal{B_{C}} \rightarrow \mathcal{B}_{n}K_{S}^{0})-\Gamma(\mathcal{B_{C}}\rightarrow\mathcal{B}_{n}K_{L}^{0})}{\Gamma(\mathcal{B_{C}}\rightarrow\mathcal{B}_{n}K_{S}^{0})+\Gamma(\mathcal{B_{C}}\rightarrow\mathcal{B}_{n}K_{L}^{0})}\label{R}\;.
\end{align}
After using relations in Eq.(\ref{relations}), the asymmetry can be further written as the following
\begin{align}
R(\mathcal{B}_{C}\rightarrow\mathcal{B} K_{S,L}^{0})=\frac{-2H_{DC}*H_{CA}}{H_{DC}^{2}+H_{CA}^{2}}\;.
\end{align}

Here $H_{CA}$ and $H_{DC}$ are respectively the Cabibbo allowed and doubly Cabibbo suppressed $SU(3)$ amplitude for the processes of final states $K_{S,L}^{0}$ mixed by $K^{0}$ and $\bar{K}^{0}$, which are shown in the Table~\ref{table1}. Using Table~\ref{parameters}, we predict the asymmetries for several processes which are shown in the last column in Table~\ref{rpre}.  The $K_{S}^{0}-K_{L}^{0}$ asymmetries could be used to search for the doubly Cabibbo suppressed charmless two body hadronic decays of anti-triplet charmed baryon decays.

\section{Conclusions}

In this work, we have carried out an analysis for charmless two body decays of anti-triplet charmed baryons based on $SU(3)$ flavour symmetry. Combined with other known data on such decays and those newly measured ones from Belle collaboration, a lot of information can be obtained, in particular making a meaningful global analysis possible. 

$SU(3)$ symmetry predicts some relations among different processes. One needs to be careful in using these relations because the predictions are sensitive to particle masses in individual decays. Using physical masses for relevant particles we find that these relations are  in good with our global analysis. For the scenario that all amplitudes are real, our global fit gives a reasonable with $\chi^2/d.o.f=0.744$. This indicates that data fit $SU(3)$ predictions with real amplitudes well.  We await more data to come to do a more general analysis. We made predictions for 29 processes for their branching fractions and polarization parameters without $\eta$ or $\eta^\prime$ decays which provide further tests for the $SU(3)$ model. 

For processes involve $\eta$ and $\eta'$ in the final states, only the class of decay with amplitudes proportional to  $a_6 - a_{15}$ have 3 decay branching fractions for data available for fitting. Even with this limitation, their polarization parameters have been predicted. We also predicted 5 decay processes with $\eta$ or $\eta^\prime$ in the final states and their associated polarization parameters to test the model.

There is another class of decays proportional to $a_6 + a_{15}$ has no data available to constraint the decays. Thus no predictions can be made. Measurements of some of thesw types of decays are crucial to test $SU(3)$ symmetry. We urge our experimental colleagues to measure some of this class of decays.
We gave several comments on the feature of global fit regarding the branching fractions, relations between different decays involving $K^0$ and $\bar K^0$. These can also be tested with future data.
We eagerly waiting for data from future experimental data from Belle, Belle~II, BESIII, and LHCb to test the validity of $SU(3)$ for charmless two body hadronic decays to better precision.
   
\section*{ACKNOWLEDGMENT}
We would like to thank Prof. Cheng-Ping Shen and Dr. Long-Ke Li for many fruitful discussions and suggestions. F.Huang and Z.P.Xing are grateful to Prof. Wei Wang and Mr. Jin Sun for inspiring discussions and valuable comments. This work was supported in part by NSFC under Grant Nos. 11735010, 11905126, U2032102, 12061131006, 12125503, 12147147, 11975149, 12090064, and  the MOST (Grant No. MOST 106-2112-M-002-003-MY3 ).


\begin{appendix}

 \begin{table}[htbp!]
\caption{SU(3) amplitudes and predicted branching fractions (the third column) and polarization parameters (the fourth column) of anti-triplet charmed baryons  decays into an octet baryon and an octet meson.}\label{table1}\begin{tabular}{|c|c|c|c|c|c|c|c}\hline\hline
channel &SU(3) amplitude& branching ratio($10^{-2}$)&$\alpha$\\\hline \hline
$\Lambda^{+}_{c}\to \Sigma^{0}  \pi^{+} $ & $ (-b_6+b_{15}+c_6-c_{15}+d_6)/\sqrt{2}$&$1.272\pm0.056$&$-0.605\pm0.088$\\\hline
$\Lambda^{+}_{c}\to \Lambda  \pi^{+} $ & $ -(b_6-b_{15}+c_6-c_{15}+d_6+2 e_{15})/\sqrt{6}$&$1.307\pm0.069$&$-0.841\pm0.083$\\\hline
$\Lambda^{+}_{c}\to \Sigma^{+}  \pi^{0} $ & $ (b_6-b_{15}-c_6+c_{15}-d_6)/\sqrt{2}$&$1.283\pm0.057$&$-0.603\pm0.088$\\\hline
$\Lambda^{+}_{c}\to p  K_{S}^{0} $ & $ ( \sin^2\theta \left(-d_6+d_{15}+e_{15}\right)+b_6-b_{15}-e_{15})/\sqrt{2}$&$1.587\pm0.077$&$0.19\pm0.41$\\\hline
$\Lambda^{+}_{c}\to \Xi^{0}  K^{+} $ & $ -c_6+c_{15}+d_{15}$&$0.548\pm0.068$&$0.866\pm0.090$\\\hline
$\Xi^{+}_{c}\to \Sigma^{+}  K_{S}^{0} $ & $  (\sin^2\theta \left(b_6-b_{15}-e_{15}\right)-d_6+d_{15}+e_{15})/\sqrt{2}$&$0.53\pm0.70$&$0.6\pm2.2$\\\hline
$\Xi^{+}_{c}\to \Xi^{0}  \pi^{+} $ & $ -d_6-d_{15}-e_{15}$&$0.54\pm0.18$&$-0.94\pm0.15$\\\hline
$\Xi^{0}_{c}\to \Sigma^{0}  K_{S}^{0} $ & $  (-\sin^2\theta \left(b_6+b_{15}-e_{15}\right)+(c_6+c_{15}+d_6-e_{15}))/2$&$0.069\pm0.024$&$1.00\pm0.13$\\\hline
\multirow{2}{*}{$\Xi^{0}_{c}\to \Lambda  K^0_S$ }  &$ \sqrt{3}\sin^2\theta \left(b_6+b_{15}-2 c_6-2 c_{15}-2 d_6+e_{15}\right)/6$&\multirow{2}{*}{$0.334\pm0.065$}&\multirow{2}{*}{$-0.04\pm0.63$}\cr
&$+ \sqrt{3}(2 b_6+2 b_{15}-c_6-c_{15}-d_6-e_{15})/6$&&\\\hline
$\Xi^{0}_{c}\to \Sigma^{+}  K^{-} $ & $ c_6+c_{15}+d_{15}$&$0.221\pm0.068$&$-0.9\pm1.0$\\\hline
$\Xi^{0}_{c}\to \Xi^{-}  \pi^{+} $ & $ b_6+b_{15}+e_{15}$&$1.21\pm0.21$&$-0.56\pm0.32$\\\hline
$\Xi^{0}_{c}\to \Xi^{0}  \pi^{0} $ & $ (-b_6-b_{15}+d_6+d_{15})/\sqrt{2}$&$0.256\pm0.093$&$-0.23\pm0.60$\\\hline
$\Lambda^{+}_{c}\to \Sigma^{0}  K^{+} $   & $ \sin\theta \left(-b_6+b_{15}+d_6+d_{15}\right)/\sqrt{2}$&$0.0504\pm0.0056$&$-0.953\pm0.040$\\\hline
$\Lambda^{+}_{c}\to \Lambda  K^{+} $  & $ -\sin\theta \left(b_6-b_{15}-2 c_6+2 c_{15}+d_6+3 d_{15}+2 e_{15}\right)/\sqrt{6}$&$0.064\pm0.010$&$-0.24\pm0.15$\\\hline
$\Lambda^{+}_{c}\to \Sigma^{+} K^0_S/K^0_L $   & $ \sin\theta \left(-b_6+b_{15}+d_6-d_{15}\right)$&$0.0103\pm0.0042$&$-0.13\pm0.98$\\\hline
$\Lambda^{+}_{c}\to p  \pi^{0} $          & $ \sin\theta \left(-c_6+c_{15}-d_6+e_{15}\right)/\sqrt{2}$&$0.445\pm0.085$&$0.663\pm0.069$\\\hline
$\Lambda^{+}_{c}\to n  \pi^{+}$           & $ -\sin\theta \left(c_6-c_{15}+d_6+e_{15}\right)$&$0.035\pm0.011$&$0.77\pm0.28$\\\hline
$\Xi^{+}_{c}\to \Sigma^{0}  \pi^{+} $     & $ -\sin\theta \left(b_6-b_{15}-c_6+c_{15}+d_{15}+e_{15}\right)/\sqrt{2}$&$0.318\pm0.027$&$-0.649\pm0.076$\\\hline
$\Xi^{+}_{c}\to \Lambda  \pi^{+} $    & $ \sin\theta \left(-b_6+b_{15}-c_6+c_{15}+2 d_6+3 d_{15}+e_{15}\right)/\sqrt{6}$&$0.056\pm0.014$&$-0.06\pm0.35$\\\hline
$\Xi^{+}_{c}\to \Sigma^{+}  \pi^{0} $     & $ \sin\theta \left(b_6-b_{15}-c_6+c_{15}-d_{15}-e_{15}\right)/\sqrt{2}$&$0.24\pm0.20$&$-0.993\pm0.091$\\\hline
$\Xi^{+}_{c}\to p  K^0_S/K^0_L $          & $ \sin\theta \left(-b_6+b_{15}+d_6-d_{15}\right)$&$0.099\pm0.083$&$-0.4\pm3.0$\\\hline
$\Xi^{+}_{c}\to \Xi^{0}  K^{+} $          & $ -\sin\theta \left(c_6-c_{15}+d_6+e_{15}\right)$&$0.09\pm0.12$&$0.99\pm0.16$\\\hline
$\Xi^{0}_{c}\to \Sigma^{0}  \pi^{0} $     & $ -\frac{1}{2} \sin\theta \left(b_6+b_{15}+c_6+c_{15}-d_{15}-e_{15}\right)$&$0.0032\pm0.0090$&$-0.98\pm0.17$\\\hline
$\Xi^{0}_{c}\to \Lambda  \pi^{0} $    & $ \sin\theta \left(b_6+b_{15}+c_6+c_{15}-2 d_6-3 d_{15}+e_{15}\right)/2 \sqrt{3}$&$0.044\pm0.012$&$0.94\pm0.15$\\\hline
$\Xi^{0}_{c}\to \Sigma^{+}  \pi^{-} $     & $ -\sin\theta \left(c_6+c_{15}+d_{15}\right)$&$0.0131\pm0.0049$&$-0.96\pm0.85$\\\hline
$\Xi^{0}_{c}\to p  K^{-} $                & $ \sin\theta \left(c_6+c_{15}+d_{15}\right)$&$0.026\pm0.078$&$-0.8\pm1.5$\\\hline
$\Xi^{0}_{c}\to \Sigma^{-}  \pi^{+} $     & $ -\sin\theta \left(b_6+b_{15}+e_{15}\right)$&$0.080\pm0.014$&$-0.50\pm0.30$\\\hline
$\Xi^{0}_{c}\to n  K^0_S/K^0_L $& $ \sin\theta \left(-b_6-b_{15}+c_6+c_{15}+d_6\right)$&$0.018\pm0.033$&$-0.98\pm0.32$\\\hline
$\Xi^{0}_{c}\to \Xi^{-}  K^{+} $          & $ \sin\theta \left(b_6+b_{15}+e_{15}\right)$&$0.0470\pm0.0083$&$-0.60\pm0.33$\\\hline
$\Xi^{0}_{c}\to \Xi^{0}  K^0_S/K^0_L $          & $ \sin\theta \left(b_6+b_{15}-c_6-c_{15}-d_6\right)$&$0.019\pm0.013$&$-0.44\pm0.67$\\\hline
$\Lambda^{+}_{c}\to p  K^0_L $ 		& $ (\sin^2\theta \left(-d_6+d_{15}+e_{15}\right)-b_6+b_{15}+e_{15})/\sqrt{2}$&$1.70\pm0.12$&$0.24\pm0.47$\\\hline
$\Lambda^{+}_{c}\to n  K^{+} $ 		& $ \sin^2\theta \left(d_6+d_{15}+e_{15}\right)$&$0.0024\pm0.0011$&$-0.41\pm0.17$\\\hline
$\Xi^{+}_{c}\to \Sigma^{0}  K^{+} $ & $ \sin^2\theta \left(b_6-b_{15}+e_{15}\right)/\sqrt{2}$&$0.01223\pm0.00057$&$-0.985\pm0.019$\\\hline
$\Xi^{+}_{c}\to \Lambda  K^{+} $& $\sin^2\theta \left(b_6-b_{15}-2 c_6+2 c_{15}-2 d_6-e_{15}\right)/\sqrt{6}$&$0.00365\pm0.00053$&$0.46\pm0.11$\\\hline
$\Xi^{+}_{c}\to \Sigma^{+}  K^0_L $ & $ (\sin^2\theta \left(b_6-b_{15}-e_{15}\right)+d_6-d_{15}-e_{15})/\sqrt{2}$&$0.97\pm0.89$&$0.4\pm1.8$\\\hline
$\Xi^{+}_{c}\to p  \pi^{0} $ 		& $\sin^2\theta \left(c_6-c_{15}+d_{15}\right)/\sqrt{2}$&$0.040\pm0.049$&$0.48\pm0.29$\\\hline
$\Xi^{+}_{c}\to n  \pi^{+} $ 		& $ \sin^2\theta \left(c_6-c_{15}-d_{15}\right)$&$0.0132\pm0.0035$&$0.70\pm0.10$\\\hline
$\Xi^{0}_{c}\to \Sigma^{0}  K_{L}^{0}$ & $  (-\sin^2\theta \left(b_6+b_{15}-e_{15}\right)-(c_6+c_{15}+d_6-e_{15}))/2$&$0.091\pm0.028$&$0.99\pm0.20$\\\hline
\multirow{2}{*}{$\Xi^{0}_{c}\to \Lambda  K^0_L$ }  &$ \sqrt{3}\sin^2\theta \left(b_6+b_{15}-2 c_6-2 c_{15}-2 d_6+e_{15}\right)/6$&\multirow{2}{*}{$0.335\pm0.065$}&\multirow{2}{*}{$-0.04\pm0.63$}\cr
&$- \sqrt{3}(2 b_6+2 b_{15}-c_6-c_{15}-d_6-e_{15})/6$&&\\\hline
$\Xi^{0}_{c}\to p  \pi^{-} $ 		& $ \sin^2\theta \left(c_6+c_{15}+d_{15}\right)$&$0.0014\pm0.043$&$-0.8\pm1.1$\\\hline
$\Xi^{0}_{c}\to \Sigma^{-}  K^{+} $ & $ \sin^2\theta \left(b_6+b_{15}+e_{15}\right)$&$0.00328\pm0.00058$&$-0.53\pm0.31$\\\hline
$\Xi^{0}_{c}\to n  \pi^{0} $ 		& $ -\sin^2\theta \left(c_6+c_{15}-d_{15}\right)/\sqrt{2}$&$0.0026\pm0.0027$&$0.80\pm0.58$\\\hline
\hline
\end{tabular}
\end{table}



\begin{table}[htbp!]
\caption{SU(3) amplitudes and  predicted branching fractions (the third column) and polarization parameters (the fourth column) of anti-triplet charmed baryons  decays into an octet baryon and  $\eta$ or $\eta^\prime$. In this table ``$-$" represent the channel can not be prediction due to the limit of experimental data.}\label{tableeta}
\begin{tabular}{|c|c|c|c|c|c|c|c}\hline\hline
channel &SU(3) amplitude& branching fraction($10^{-2}$)&$\alpha$\\\hline \hline
$\Lambda^{+}_{c}\to \Sigma^{+}  \eta $ & $\cos\phi(-2 a_6+2 a_{15}-b_6+b_{15}-c_6+c_{15}+d_6)/\sqrt{2}-\sin\phi( -a_6+a_{15}+d_{15})$&$0.45\pm0.19$&$0.3\pm3.8$\\\hline
$\Lambda^{+}_{c}\to \Sigma^{+}  \eta^\prime $ & $\sin\phi(-2 a_6+2 a_{15}-b_6+b_{15}-c_6+c_{15}+d_6)/\sqrt{2}+\cos\phi( -a_6+a_{15}+d_{15})$&$1.5\pm0.6$&$0.8\pm1.9$\\\hline
\multirow{2}{*}{$\Lambda^{+}_{c}\to p  \eta $ }       & $\sin\theta\big(\cos\phi \left(-2 a_6+2 a_{15}-c_6+c_{15}+d_6-e_{15}\right)/\sqrt{2}$&\multirow{2}{*}{$0.127\pm0.024$}&\multirow{2}{*}{$-0.02\pm4.93$}\cr
&$- \sin\phi \left(-a_6+a_{15}-b_6+b_{15}+d_{15}+e_{15}\right)\big)$&&\\\hline
\multirow{2}{*}{$\Lambda^{+}_{c}\to p  \eta^\prime $}         & $\sin\theta\big(\sin\phi \left(-2 a_6+2 a_{15}-c_6+c_{15}+d_6-e_{15}\right)/\sqrt{2}$&\multirow{2}{*}{$0.27\pm0.38$}&\multirow{2}{*}{$-0.2\pm9.0$}\cr
&$+ \cos\phi \left(-a_6+a_{15}-b_6+b_{15}+d_{15}+e_{15}\right)\big)$&&\\\hline
\multirow{2}{*}{$\Xi^{+}_{c}\to \Sigma^{+} \eta $ }   & $\sin\theta \big(\cos\phi\left(-2 a_6+2 a_{15}-b_6+b_{15}-c_6+c_{15}+d_{15}+e_{15}\right)/\sqrt{2} $&\multirow{2}{*}{$0.12\pm0.13$}&\multirow{2}{*}{$-0.2\pm2.6$}\cr
&$-\sin\phi \left(-a_6+a_{15}+d_6-e_{15}\right)\big)$&&\\\hline
\multirow{2}{*}{$\Xi^{+}_{c}\to \Sigma^{+}  \eta^\prime $ }   &$\sin\theta \big(\sin\phi\left(-2 a_6+2 a_{15}-b_6+b_{15}-c_6+c_{15}+d_{15}+e_{15}\right)/\sqrt{2} $&\multirow{2}{*}{$0.184\pm0.083$}&\multirow{2}{*}{$0.5\pm4.6$}\cr
&$+\cos\phi \left(-a_6+a_{15}+d_6-e_{15}\right)\big)$&&\\\hline
$\Xi^{+}_{c}\to p  \eta $ 		& $ \sin^2\theta (\cos\phi\left(2 a_6-2 a_{15}+c_6-c_{15}-d_{15}\right)-\sin\phi\left(a_6-a_{15}+b_6-b_{15}-d_6\right))$&$0.05\pm0.12$&$-0.22\pm0.11$\\\hline
$\Xi^{+}_{c}\to p  \eta^\prime $ 		& $ \sin^2\theta (\cos\phi\left(2 a_6-2 a_{15}+c_6-c_{15}-d_{15}\right)+\sin\phi\left(a_6-a_{15}+b_6-b_{15}-d_6\right))$&$0.07\pm0.43$&$0.6\pm1.9$\\\hline
$\Xi^{0}_{c}\to \Xi^{0}  \eta $ & $ \cos\phi(2 a_6+2 a_{15}+b_6+b_{15}-d_6+d_{15})/\sqrt{2}-\sin\phi(a_6+a_{15}+c_6+c_{15})$&-&-\\\hline
$\Xi^{0}_{c}\to \Xi^{0}  \eta^\prime $ & $ \sin\phi(2 a_6+2 a_{15}+b_6+b_{15}-d_6+d_{15})/\sqrt{2}+\cos\phi(a_6+a_{15}+c_6+c_{15})$&-&-\\\hline
\multirow{2}{*}{$\Xi^{0}_{c}\to \Sigma^{0}  \eta $}    & $ \sin\theta \big(\
cos\phi\left(2 a_6+2 a_{15}+b_6+b_{15}+c_6+c_{15}+d_{15}-e_{15}\right)/2$&\multirow{2}{*}{-}&\multirow{2}{*}{-}\cr
&$-\sin\phi \left(a_6+a_{15}-d_6+e_{15}\right)/\sqrt{2}\big)$&&\\\hline
\multirow{2}{*}{$\Xi^{0}_{c}\to \Sigma^{0}  \eta^\prime$}    & $ \sin\theta (\sin\phi\left(2 a_6+2 a_{15}+b_6+b_{15}+c_6+c_{15}+d_{15}-e_{15}\right)/2$&\multirow{2}{*}{-}&\multirow{2}{*}{-}\cr
&$-\cos\phi \left(a_6+a_{15}-d_6+e_{15}\right)/\sqrt{2}\big)$&&\\\hline
\multirow{2}{*}{$\Xi^{0}_{c}\to \Lambda  \eta $}   & $ \sin\theta \big( -\cos\phi \left(6 a_6+6 a_{15}+b_6+b_{15}+c_6+c_{15}-2 d_6+3 d_{15}+e_{15}\right)/(2 \sqrt{3})$&\multirow{2}{*}{-}&\multirow{2}{*}{-}\cr
&$-\sin\phi\left(-3 a_6-3 a_{15}-2 b_6-2 b_{15}-2 c_6-2 c_{15}+d_6+e_{15}\right)/\sqrt{6}\big)$&&\\\hline
\multirow{2}{*}{$\Xi^{0}_{c}\to \Lambda  \eta^\prime$ }  &$ \sin\theta \big( -\sin\phi \left(6 a_6+6 a_{15}+b_6+b_{15}+c_6+c_{15}-2 d_6+3 d_{15}+e_{15}\right)/(2 \sqrt{3})$&\multirow{2}{*}{-}&\multirow{2}{*}{-}\cr
&$+\cos\phi\left(-3 a_6-3 a_{15}-2 b_6-2 b_{15}-2 c_6-2 c_{15}+d_6+e_{15}\right)/\sqrt{6}\big)$&&\\\hline
\multirow{2}{*}{$\Xi^{0}_{c}\to n \eta $ }       & $\sin^2\theta(\cos\phi\left(2 a_6+2 a_{15}+c_6+c_{15}+d_{15}\right)/\sqrt{2}$&\multirow{2}{*}{-}&\multirow{2}{*}{-}\cr
&$- \sin\phi \left(a_6+a_{15}+b_6+b_{15}-d_6\right)\big)$&&\\\hline
\multirow{2}{*}{$\Xi^{0}_{c}\to n \eta $ }       & $\sin^2\theta(\sin\phi\left(2 a_6+2 a_{15}+c_6+c_{15}+d_{15}\right)/\sqrt{2}$&\multirow{2}{*}{-}&\multirow{2}{*}{-}\cr
&$+\cos\phi  \left(a_6+a_{15}+b_6+b_{15}-d_6\right)\big)$&&\\\hline
\hline
\end{tabular}
\end{table}

\begin{table}[htbp!]
\caption{Experimental data on branching fractions and polarization parameters (the second column), previous theoretical predictions (the third column) and our fitting results (the fourth column).}\label{exp1}
\begin{tabular}{|c|c|c|c|c|c|c|c|c|c|}\hline\hline
\multirow{2}{*}{channel} &\multicolumn{5}{c|}{ branching fraction}\cr\cline{2-6}
&Experimental data($10^{-2}$)&\multicolumn{3}{c|}{SU(3) symmetry analysis ($10^{-2}$)}&Our work ($10^{-2}$)\\\hline
$\Lambda^{+}_{c}\to p K_S^0 $ & $1.59\pm0.08$ &$0.61\pm0.07$\cite{Geng:2019xbo}&$1.46\pm0.47$\cite{Zhao:2018mov}&$1.36\sim 1.80$\cite{Wang:2017gxe}& $1.587\pm0.077$ \\\hline
$\Lambda^{+}_{c}\to p\eta $ &$0.124\pm0.03$&$0.124\pm0.035$\cite{Geng:2019xbo}&$0.114\pm0.035$\cite{Zhao:2018mov}&-&$0.127\pm0.024$\\\hline
$\Lambda^{+}_{c}\to \Lambda \pi^+$ & $1.3\pm0.07$&$1.3\pm0.07$\cite{Geng:2019xbo}&$1.32\pm0.34$\cite{Zhao:2018mov}&$1.30\pm0.17$\cite{Wang:2017gxe}&$1.307\pm0.069$\\\hline
$\Lambda^{+}_{c}\to \Sigma^0\pi^+ $ & $1.29\pm0.07$&$1.27\pm0.06$\cite{Geng:2019xbo}&$1.26\pm0.32$\cite{Zhao:2018mov}&$1.27\pm0.17$\cite{Wang:2017gxe}&$1.272\pm0.056$\\\hline
$\Lambda^{+}_{c}\to \Sigma^{+}\pi^0 $ & $1.25\pm0.10$&$1.27\pm0.06$\cite{Geng:2019xbo}&$1.23\pm0.17$\cite{Zhao:2018mov}&$1.27\pm0.17$\cite{Wang:2017gxe}&$1.283\pm0.057$\\\hline
$\Lambda^{+}_{c}\to \Xi^{0}K^{+} $ & $0.55\pm0.07$&$0.56\pm0.09$\cite{Geng:2019xbo}&$0.59\pm0.17$\cite{Zhao:2018mov}&$0.50\pm0.12$\cite{Wang:2017gxe}&$0.548\pm0.068$\\\hline
$\Lambda^{+}_{c}\to \Lambda K^{+} $ & $0.061\pm0.012$&$0.065\pm0.010$\cite{Geng:2019xbo}&$0.059\pm0.017$\cite{Zhao:2018mov}&-&$0.064\pm0.010$\\\hline
$\Lambda^{+}_{c}\to \Sigma^{+}\eta $ & $0.44\pm0.20$&$0.32\pm0.31$\cite{Geng:2019xbo}&$0.47\pm0.22$\cite{Zhao:2018mov}&-&$0.45\pm0.19$\\\hline
$\Lambda^{+}_{c}\to \Sigma^{+}\eta^\prime $ & $1.5\pm0.60$&$1.44\pm0.56$\cite{Geng:2019xbo}&$0.93\pm0.28$\cite{Zhao:2018mov}&-&$1.5\pm0.60$\\\hline
$\Lambda^{+}_{c}\to \Sigma^{0}K^+ $ & $0.052\pm0.008$&$0.054\pm0.007$\cite{Geng:2019xbo}&$0.055\pm0.016$\cite{Zhao:2018mov}&-&$0.0504\pm0.0056$\\\hline
$\Xi^{+}_{c}\to \Xi^{0}\pi^+ $ & $1.6\pm0.8$&$3.8\pm2.0$\cite{Geng:2019xbo}&$0.93\pm0.36$\cite{Zhao:2018mov}&$0.01\sim10.22$\cite{Wang:2017gxe}&$ 0.54\pm0.18$\\\hline
$\Xi^{0}_{c}\to \Lambda K_S^0 $ & $0.334\pm0.067$& $5.25\pm0.3$\cite{Geng:2019xbo}&$4.15\pm2.5$\cite{Zhao:2018mov}&$0.47\pm0.08$\cite{Wang:2017gxe}&$0.334\pm0.065$\\\hline
$\Xi^{0}_{c}\to \Xi^- \pi^+ $ & $1.43\pm0.32$&$2.21\pm0.14$\cite{Geng:2019xbo}&$0.37\pm0.22$\cite{Zhao:2018mov}&$2.24\pm0.34$\cite{Wang:2017gxe}&$1.21\pm0.21$\\\hline
$\Xi^{0}_{c}\to \Xi^- K^+ $ & $0.039\pm0.012$&$0.098\pm0.006$\cite{Geng:2019xbo}&$0.056\pm0.008$\cite{Zhao:2018mov}&-&$0.047\pm0.0083$\\\hline
$\Xi_c^0\to\Sigma^0 K^0_S$ & $0.069\pm0.024$&$0.4\pm0.4$\cite{Geng:2019xbo}&$3.95\pm2.4$\cite{Zhao:2018mov}&$0.23\pm0.07$\cite{Wang:2017gxe}&$0.069\pm0.024$\\\hline
$\Xi_c^0\to\Sigma^+ K^-$ & $0.221\pm0.068$&$5.9\pm1.1$\cite{Geng:2019xbo}&$22.0\pm5.7$\cite{Zhao:2018mov}&$3.1\pm0.9$\cite{Wang:2017gxe}&$0.221\pm0.068$\\\hline
\multirow{2}{*}{channel} &\multicolumn{5}{c|}{ asymmetry parameter $\alpha$}\cr\cline{2-6}
&Experimental data&\multicolumn{3}{c|}{SU(3) symmetry analysis} &Our work\\\hline
$\alpha(\Lambda^{+}_{c}\to p K_S^0)$&$0.18\pm0.45$&-&-&-&$0.19\pm0.41$\\\hline
$\alpha(\Lambda^{+}_{c}\to \Lambda \pi^+)$&$-0.84\pm0.09$&$-0.87\pm0.10$\cite{Geng:2019xbo}&-&-&$-0.841\pm0.083$\\\hline
$\alpha(\Lambda^{+}_{c}\to \Sigma^{0}\pi^+)$&$-0.73\pm0.18$&$-0.35\pm0.27$\cite{Geng:2019xbo}&-&-&$-0.605\pm0.088$\\\hline
$\alpha(\Lambda^{+}_{c}\to  \Sigma^{+}\pi^0)$&$-0.55\pm0.11$&$-0.35\pm0.27$\cite{Geng:2019xbo}&-&-&$-0.603\pm0.088$\\\hline
$\alpha(\Xi^{0}_{c}\to \Xi^{-} \pi^{+})$&$-0.6\pm0.4$&$-0.98^{+0.07}_{-0.02}$\cite{Geng:2019xbo}&-&-&$-0.56\pm0.32$\\\hline
$\chi^{2}/d.o.f$&-&-&-&-&0.744\\\hline
\end{tabular}
\end{table}

\begin{table}[htbp!]
\caption{SU(3) symmetry irreduciable amplitudes from fitting.}\label{data}\label{parameters}
\begin{tabular}{|c|c|c|c|c|c|c|c|c|c|c|}\hline\hline
\multirow{2}{*}{parameters}&\multicolumn{2}{c|}{ Our work }\cr\cline{2-3}
&scalar(f) & pseudoscalar(g)\\\hline
$b_{6}$  & $-0.111\pm0.0093$&$0.142\pm0.026$\\\hline
$c_{6}$  & $-0.010\pm0.018$&$-0.106\pm0.078$\\\hline
$d_{6}$  & $ -0.042\pm0.015$&$0.02\pm0.12$\\\hline
$b_{15}$  & $ 0.0448\pm0.0091$&$-0.021\pm0.019$\\\hline
$c_{15}$  & $ 0.063\pm0.018$&$0.140\pm0.052$\\\hline
$d_{15}$  & $ -0.018\pm0.014$&$-0.11\pm0.12$\\\hline
$e_{15}$  & $ 0.0382\pm0.0044$&$0.185\pm0.024$\\\hline
$a$ & $0.121\pm0.064$&$0.22\pm0.77$\\\hline
$a^\prime$  &-&-\\\hline
$\chi^{2}$/d.o.f&\multicolumn{2}{c|}{ 0.744} \cr\cline{1-3}
\end{tabular}
\end{table}

\begin{table}[htbp!]
\caption{ $K_{S}^{0}-K_{L}^{0}$ asymmetries. }\label{rpre}
\begin{tabular}{|c|c|c|c|c|c|c|c|c|c|c|}\hline\hline
Channel&$H_{CA}$&$H_{DC}$& R\\\hline
$\Lambda^{+}_{c}\to p K_{S,L}^{0} $ &$-b_{6}+b_{15}+e_{15}$&$\sin^2\theta(-d_{6}+d_{15}+e_{15})$& $0.88\pm0.09$\\\hline
$\Xi^{+}_{c}\to \Sigma^{+} K_{S,L}^{0} $ &$d_{6}-d_{15}-e_{15}$&$\sin^2\theta(b_{6}-b_{15}-e_{15})$& $-0.3\pm0.8$\\\hline
$\Xi^{0}_{c}\to \Sigma^{0} K_{S,L}^{0} $ &$ -({c_{6}+c_{15}+d_{6}-e_{15}})/\sqrt{2}$&$-(\sin^2\theta(b_{6}+b_{15}-e_{15}))/\sqrt{2}$&$-0.14\pm0.23$\\\hline
$\Lambda^{+}_{c}\to \Lambda  K_{S,L}^{0} $ &$(-2b_{6}-2b_{15}+c_{6}+c_{15}+d_{6}+e_{15})/\sqrt{6}$&$(\sin^2\theta(b_{6}+b_{15}-2c_{6}-2c_{15}-2d_{6}+e_{15})/\sqrt{6}$& $ -0.0014\pm0.137$\\\hline
\end{tabular}
\end{table}


\end{appendix}

\end{document}